\documentclass[pra,twocolumn]{revtex4}
\usepackage{amsmath,amssymb,bm,txfonts,amssymb,epsfig,graphics}
\usepackage{CJK}

\begin{document}

\begin{CJK*}{GBK}{kai}


\title{Demonstration of quantum Zeno effect in a superconducting phase qubit}

\author{Zhen-Tao Zhang$^{1,2}$, Zheng-Yuan Xue$^1$\thanks{e-mail: xuezhengyuan@yahoo.com.cn}}

\address{$^1$Laboratory of Quantum Information Technology, School
of Physics and Telecommunication Engineering,\\ South China Normal
University, Guangzhou 510006,  China\\
$^2$ National Laboratory of Solid State Microstructures, School of
Physics, Nanjing University , Nanjing 210093 , China }

\date{\today}



\begin{abstract}

Quantum Zeno effect is a significant tool in quantum manipulating
and computing. We propose its observation in superconducting phase
qubit with two experimentally feasible measurement schemes. The
conventional measurement method is used to achieve the proposed
pulse and continuous readout of the qubit state, which are analyzed
by projection assumption and Monte Carlo wave-function simulation,
respectively. Our scheme gives a direct implementation of quantum
Zeno effect in a superconducting phase qubit.

\end{abstract}
\maketitle


Quantum Zeno effect (QZE), proposed by Misra and Sudarshan in 1977
\cite{Misra}, is a paradigm showing that quantum physics is
counter-intuitive. It predict that if the state of a unstable or
oscillating quantum system is measured frequently to see  whether it
still stay at a initial state, transitions from the initial state to
other states will be suppressed or even inhibited.   Since then,
many exciting progresses have been made both theoretically and
experimentally.  In the theoretical side, physicist interpret it
with wave-function collapse assumption in early days \cite{Cook,
Itano}, which is shown to be not necessary \cite{Frerichs}. Later,
it was generalized in a few different ways. Concerning the
measurement, it can not only retard incoherent decay but also
coherent Rabi oscillation. On the other hand, for unstable system,
frequent measurements may even also enhance the decay rate under
some conditions, which is the so-called quantum anti-Zeno effect
\cite{Fischer,Kofman, Facchi}. As to the readout aspect, one can
adopt pulse or continuous measurements \cite{L. S. Schulman,Streed}.
In the experimental side, QZE have been demonstrated in many
systems, such as trapped ions \cite{Itano}, optical lattice
\cite{Fischer}, Bose-Einstein condensate \cite{Streed}, microwave
cavity \cite{Bernu}, etc.

Studying QZE is very important. Beyond the interest of fundamental
physics, it has many practical applications. These includes reducing
decoherence in quantum computing \cite{Facchi05,Franson,Hosten},
efficient preservation of spin polarized gases  \cite{Nagels},
keeping system stay in object subspace \cite{Barenco}. There are
interesting explorations of applications in superconducting qubit
systems, e.g., generation of entangled state \cite{X.-B. Wang} and
implementation of quantum switch \cite{Zhou}. Recently, the
possibility of observing QZE in  superconducting qubits is proposed
\cite{Lizuain,Matsuzaki}. However, demonstration of QZE in
superconducting system is very difficult because of the lacking of
competent measurement method. Conventionally, it was observed with
quantum non-demolition (QND) readout.   In  circuit Quantum
Electrodynamics (CQED), state of the qubit could be imprinted on the
cavity field state in a QND readout, but the signal-to-noise ratio
is so low that we must repeat considerable times to complete the QND
readout. Thus, a recent experiment \cite{A. Palacios-Laloy}
demonstrate QZE qualitatively in CQED can not guarantee its
practical applications due to the noise. Here, we suggest
experimental feasible schemes to demonstrate QZE in a
superconducting phase qubit with both pulse and continuous
measurement strategy  instead of QND readout.

Since the quadratic decay behavior (prerequisite for QZE) in the
initial decay stage of the qubit excited state has not been observed
yet, suppressing of energy relaxation in superconducting qubit is
not accessible technically. Thus, we here focus on another case of
QZE, i.e., suppressing the unitary evolution of the phase qubit.
There are at least two feathers that differentiate our schemes from
those implemented in other systems. Firstly, the measurement method
used here is the so called selective measurement instead of QND
readout which is still a big challenge to realize continuous
measurement in superconducting qubits system. Secondly, our schemes
are immune from the relaxation of the qubit by using an appropriate
initial state. That is rather necessary in the context of the very
short energy relaxation time of superconducting qubits. It should be
noticed that, besides the function of demonstrating the basic
phenomenon of quantum mechanics, our proposal can lay a foundation
for the applications of QZE in quantum information processing, e.g.,
Ref. \cite{X.-B. Wang}.



\begin{figure}\centering
\includegraphics [width=7cm]{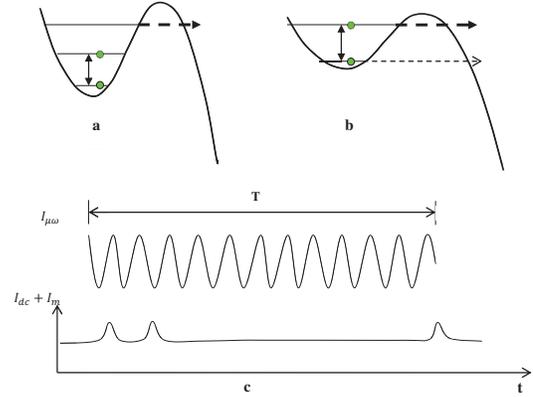}
\caption{Energy levels in the potential well when
measurement pulses is off (\textbf{a}) and on (\textbf{b});
\textbf{c}, indication of the bias of the phase qubit, including
microwave driving $I_{\mu w}$ (upper) and dc bias current with
measurement pulses (down).}
\end{figure}

Superconducting phase qubit usually consists of a large
current-biased Josephson junction (JJ). When the bias current
approaches its critical current, there exist several no-degenerate
energy levels in each well of the washboard potential of the qubit,
see Fig. 1(a). The lowest two states act as a qubit, which is the
so-called phase qubit. Experimentally, it can be easily controlled
by bias cunrrent containing both dc and microwave components
\cite{Martinis}.

An advantage of phase qubit over other types of superconducting
qubit is its built-in readout. It relies on the possibility that the
qubit states in the potential can tunnel through the potential
barrier into continuum outside. The tunneling rate of one level
usually differ dramatically from the other one at least two orders
of magnitude \cite{Martinis09}. So the ground and excited state can
be mapped to no-tunneling and tunneling case, respectively.
Experimentally, one can  lower the barrier so that the excited state
of the qubit can tunnel through the barrier quickly but the ground
state can't, see Fig. 1(b). Therefore, one can add a pulse to the
bias \cite{Cooper}, see Fig. 1(c), so that the height of the barrier
only fitting the excited qubit state.


Now, we begin our pulse measurements scheme. When operating the
phase qubit for quantum gates, the bias is tuned to an appropriate
value so that there are three or four levels in the potential well
as shown in Fig. 1(a). In this case, neither of the qubit states
could tunnel outside. Drive the phase qubit with a resonant
microwave, the qubit will oscillate between the ground and excited
state with a period of $2\pi/\Omega$. If the initial state is the
ground state, after half a period, the qubit is driven to the
excited state. To demonstrate QZE, we superpose a series of short
uniform measurement pulses to the qubit bias, see Fig1. (c). Because
the pulse during time $\tau$ is much smaller than the oscillating
period, we consider that the probe pulse is instantaneous. To be
more specifically, in half a period of Rabi oscillation
$T=\pi/\Omega$, there are $n$ evenly distributed pulses with the
time interval as $\delta t =T/n$.

According to the Hamiltonian, we can calculate straightforwardly the
population of the ground state $|0\rangle$ at $t=\delta t$, before
the first pulse, as $ P_0^-(\delta t)=\cos^2\left(\pi/2n\right).$
After the first probe, the probability of no-tunneling $P_0^1$, i.e,
the probability of the qubit collapsing to ground state, equals to
$P_0^-(\delta t)$. So, after all the $n$th probes, the survival
probability of the initial ground state $|0\rangle$ is
$P_0^n=\left(P_0^1\right)^n=\left[\cos^2\left(\pi/2n\right)\right]^n.$
When $n\gg 1$, making the proximation of
$\cos\left(\frac{\pi}{2n}\right)\approx 1-\left(\pi/2n\right)^2$ and
using the relation
$\underset{n\rightarrow\infty}\lim(1-x/n)^n=e^{-x},$ one gets
\begin{eqnarray}
P_0^n=\exp\left({-{{\pi}^2\over 4n}}\right).
\end{eqnarray}
We have plotted the survival probability with both expressions in
Fig. 2, which shows that the approximation to exponential function
is perfect when $n>10$. Obviously, with the increasing of $n$,  the
survival possibility of initial state tends to 1, which is one kind
of QZE.

\begin{figure}\centering
  \includegraphics [width=7cm]{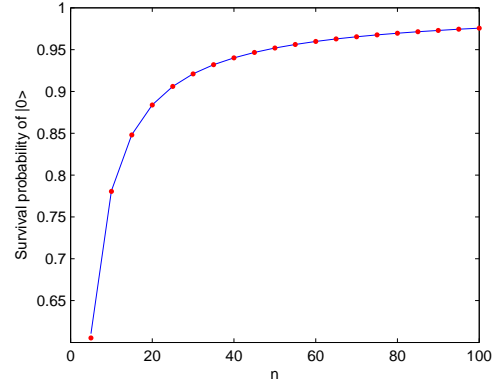}
\caption{ Survival probability of $|0\rangle$ in a half
period of Rabi oscillation vs. number of probes. Red dots is plotted
with $P_0^n=\left[\cos^2(\pi/2n)\right]^n$, and solid line with
$P_0^n=\exp({-{\pi}^2/4n})$. The approximation to
  exponential function is perfect when $n>10$.}
  \end{figure}

Although it is similar with the experiment with trapped ions
\cite{Itano}, there has significant difference between them. In
their experiment, they use a series of QND measurements during the
Rabi oscillation and measure the final state of the trapped ions at
the end of a half oscillating period. Therefore, even if some
measurement results are not the initial state, the qubit still
evolve according to Hamiltonian after the measurements. Thus, there
exist a small probability to return to the initial state at last.
Instead, we employ a selective measurement approach to obtain the
probability that  all the probes get the same result, i.e, the
initial ground state. This is achieved by the fact that if the
result is other than the initial state, the JJ will switch to a
non-zero voltage state, which means that the state of qubit will be
destructed and stop evolve after the probe. It should be noted that
our scheme exhibit what Misra and Sudarshan first called QZE
\cite{Misra}.

\begin{figure}\centering
\includegraphics [width=7cm]{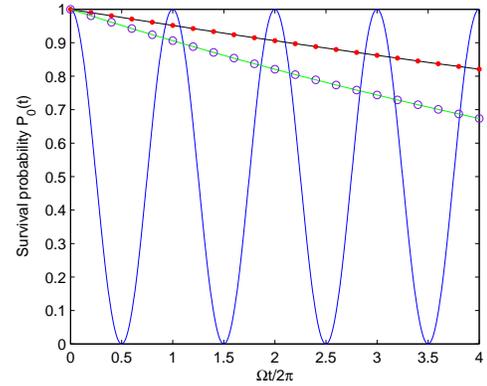}
\caption{ Survival probability of $|0\rangle$ vs. time. Blue
solid line is showing Rabi oscillation without any decay. Green and
black line is plotted according to $\cos^{2n}(\Omega\delta t/2)$
with $\delta t=\pi/50\Omega$, $\pi/100\Omega$, respectively. Purple
circle and red dot denote $\exp((-\Omega^2\delta t/4)t)$ with
corresponding $\delta t$ as above. }
\end{figure}

It is well known that when the interval time among the measurement
pulses is very small, the survival probability of the initial state
reduces exponentially with the increasing of time. Similarly, we can
easily get
 \begin{equation}
P_0(t)=p_0^n\approx\exp\left[\left(-{\Omega^2\delta t\over
4}\right)t\right]
\end{equation}
with $t=n\delta t$. We have plotted $P_0(t)$ with $\delta
t=\pi/50\Omega$, $\pi/100\Omega$ in Fig. 3. Instead of normal
Rabi-type oscillation, the initial state $|0\rangle$ decay
exponentially with an effective characteristic time $t_c$
  \begin{equation}\label{Tc}
{1\over t_c}={\Omega^2\delta t\over 4}.
  \end{equation}
It is a very important characteristic parameter showing that to what
extent the probes suppress the state transition. When $\delta
t=\pi/50\Omega$, $t_c=200/\pi\Omega$, which is much larger than the
characteristic time of Rabi oscillation $1/\Omega$,  we can also see
from Eq. (\ref{Tc}) that the decay time is inverse function of
$\delta t$.


Then, we move to the feasibility of the above pulse measurements
scheme. Theoretically, the more frequently a qubit state is
observed, the more likely it will be inhibit to transition to other
state from the initial state. However, in practice,  there are three
main obstacles stopping us from beating the QZE limit. Firstly,
measurement fidelity is always lower than unit; secondly, each
measurement is not instantaneous but inevitably lasts for a finite
period of time; and finally, the finite decoherence time of the
qubit.

The measurement of phase qubit state is achieved  by using its
macroscopic quantum tunnel. The imperfect fidelity is induced from
the finite ratio of the tunneling rates of $|1\rangle$ and
$|0\rangle$ states. It's believed that the ratio is typically around
$200$ \cite{Martinis09}. During the measurement pulse,  the
tunneling possibility from the excited state is close but a little
bit lower than 1, while that of the ground state is a small but
nonzero quantity. However, this measurement have single shot readout
fidelity up to 96\% theoretically \cite{Martinis09}, which is the
highest among all the known readout approaches of superconducting
qubits.

The other bothering factor is that the readout can't be accomplished
instantaneously. In a measurement process, the added probe pulse
alters the level structure of the qubit, making it detuning from the
driving microwave. The question how to judge the suppression of the
oscillation comes from QZE or from the reduce of the effective
driving time is unavoidable to any experimental scheme of
demonstrating QZE. Actually, for larger measuremnt times $n$, the
sum of the measurement periods is not negligible compared to the
duration of the whole process. Therefore, part of the decrease in
the transition probability is due to the decrease in the time during
which the qubit is resonant driven. For an extreme case of $n=100$,
the sum of the measurement periods is 50\% of the total time $T$.
Even for this case, the survival probability of the initial state is
as high as 97\%, which is much higher than that of sole resonant
driven.  So we can safely conclude that the suppression of the
oscillation mainly comes from QZE.

Finally, the decoherence including relaxation and pure dephasing of
the qubit is usually considered as  bottle-neck for illustrating
QZE. From now on, we would clarify why it can be neglected in our
scheme. On the one hand, pure dephasing can be ignored after
noticing that the interval between two nearest measurements is very
short compared with the dephasing time which could be as long as
hundreds of $ns$. On the other hand, the lifetime of an average
phase qubit is $T_1=600$ n$s$. If $T$ can be much smaller than
$T_1$, to avoid the decay of the excited state, then the experiment
will be able to carried out easily. But that is not necessary for
observing QZE. The reason is that before each probe, the average
population of excited state is much lower than 1, i.e.,
\begin{equation}
P_1=\frac{1}{\tau}\int_0^\tau \sin^2(\Omega
t/2)dt=\Omega^2\tau^2/12\ll 1.
\end{equation}
To be more specifically, $P_1\simeq0.07$ for $n=32$, i.e., each
probe will project the qubit sate to the ground sate subspace with
high probability. Therefore, the qubit state can only has a small
probability to excite to the excited sate, which means we will have
a much longer effective lifetime for the excited state in our
scheme. If we conservatively choose $T=T_1=600$ n$s$ and each probe
pulse lasts $t_p=3$ n$s$, then the ratio between the measurement
time and the total oscillating time is $T/t_p=200$. In the trapped
ions experiment, the quantity is about 100. With a larger ratio, one
can implement more probes within a half period of Rabi oscillation.
So, we strongly believe that QZE can be verified definitely in phase
qubit with our pulse measurement method.


Next, we propose to demonstrate QZE in a phase qubit by continuous
measurement. Theoretically, Heisenberg uncertainty principal limits
how frequently a measurement can be performed. However, one could
also adopt continuous measurement approach to observe QZE.  The main
difference of the continuous measurement approach from the above
pulse measurement scheme is that the bias is fixed to only allow the
excited state to tunnel outside during the oscillation of the qubit.
This is reasonable since the tunneling rate of the excite state is
two orders of magnitude larger than that of the ground
state\cite{Martinis09}. Furthermore, the tunneling rate of the
ground state is also smaller than the Rabi frequency of the qubit in
our parameter figuration. Therefore, we do not need to take the rare
tunneling event of the ground state into account. The system is
initially prepared in the ground state; a resonate microwave is
driving the phase qubit between the ground and excited state. The
life time of the excited state due to spontaneous decay is much
longer because of quantum tunneling is very fast. Neglecting the
spontaneous decay term, we get a Hamiltonian describing this
dissipative system in interaction picture is
\begin{equation} \label{ha2}
 {{H}_{I}}=\left( \begin{matrix}
    0 & \Omega   \\
\   \Omega  & -i\Gamma /2
 \end{matrix} \right).
\end{equation}

Before going into QZE, we would like to discuss how continue
measurement works. In our case, the excited state has a tunneling
rate $\Gamma$, but the ground state can't tunnel, which means in a
short time interval $\delta t$, the excited state tunnel with the
possibility of $\Gamma\delta t$. If a tunneling count, we know the
qubit state before tunneling is the excited state; otherwise we
can't discern the ground and excite state, but what we can get from
the interrogation is that the qubit is more likely in ground state
at the end of $\delta t$ than at the beginning. This point is the
the essential of Monte Carlo wave-function method, which is
developed for simulating open system \cite{Blatt,Molmer,Yu-Wen}.
Below we use this method to show QZE and compare it to the
analytical result.

Back to the Hamiltonian in  Eq. (\ref{ha2}).  If $\Gamma=0$, the
qubit is absolutely populates $|1\rangle$ at $t=T=\pi/2\Omega$. Now,
we suppose  $\Omega\ll\Gamma$. The non-Hermitian Hamiltonian can be
simulated with Monte Carlo wave-function method. The procedure can
be summarized as follows: (1). Discretize the time interval $T$ by a
very small time step $\delta t$. (2). Determine the probability of
tunneling $ P=\Gamma\delta t|\langle 1|\psi\rangle|^2 $, choosing
$\delta t$ to make sure $P\ll1$. (3). Obtain a random number $r$
distributed uniformly between zero and one, and compare it with $P$.
(4). If $r<P$, there is a tunneling, the system switch to finite
voltage state, and this run is end. Then start next run  from step
1. If $r>p$, no tunneling takes place, the qubit evolves under the
influence of the non-Hermitian Hamiltonian described by
Eq.(\ref{ha2}) and the qubit state at the end of $\delta t$ is
 \begin{equation}
 |\psi(t+\delta t)\rangle=(1-\frac{iH\delta
 t}{\hbar})|\psi(t)\rangle/||(1-\frac{iH\delta
 t}{\hbar})|\psi(t)\rangle||
 \end{equation}
where we have approximately expand the the evolution operator to
first order of $\delta t$. (5). Repeating the process, we can get a
trajectory of the qubit state.

It is obvious that if no tunneling appears in the whole period $T$,
the qubit state will totally stay in the state of $|0\rangle$ at the
end of the probed oscillation. So, the survival probability of a
initial state is same as that of no tunneling during the period. We
simulate many times for a certain $\Gamma$ to obtain survival
probability $P_0$. Furthermore, we have studied the relation between
$P_0$ and $\Gamma$ as shown in Fig. 4.

\begin{figure}\centering
\includegraphics [width=7cm]{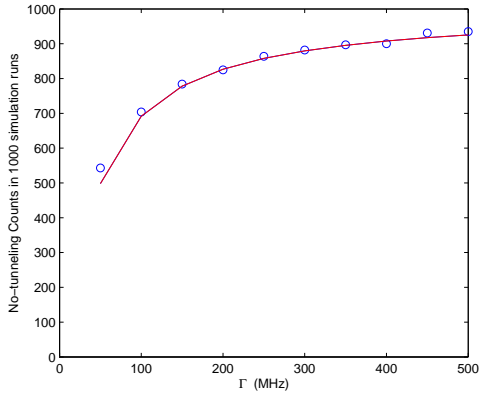}
\caption{No-tunneling counts vs tunneling rate of the
excited state. Monte Carlo wave-function simulation parameters are
$\Omega=2\pi\times1$ MHz, $T=\pi/2\Omega$ and $\Gamma \in [50, 500]$
MHz. The blue circles are data from simulation, and the red solid
line is plotted according to the analytical expression in Eq.
(\ref{A}).}
\end{figure}

The Hamiltonian in Eq. (\ref{ha2}) can also be solved analytically.
The evolution operator of the dissipative two-level system has the
form of
\begin{equation}
U=e^{-iH_It}e^{-\frac{\Gamma}{4}t}\left[\cosh(ht)-i\frac{\bm{h}\cdot
\bm{\sigma}}{h}\sinh(ht)\right]
\end{equation}
where $h=\sqrt{(\Gamma/4)^2-\Omega^2}$ and we have assumed that
$\Gamma/4>\Omega$. If the initial state is $|0\rangle$, then the
survival amplitude is function of time and has the form of
\begin{eqnarray} \label{A}
A_0(t)&=&\langle0|e^{-iH_It}|0\rangle\nonumber\\&=&e^{-\frac{\Gamma}{4}t}\left[\cosh(h
t)+\frac{\Gamma}{2h}\sinh(h t)\right].
\end{eqnarray}
We have also plotted the survival probability of the initial state
with this analytical expression in Fig. 4. We can see the Monte
Carlo wave-function simulation is agree with the analytical result
perfectly. More importantly, they both imply QZE as explained in the
following. With the increasing of the excited state tunneling rate,
the no-tunneling counts approaching to 1000, which is the total
simulation runs' number. We conclude that the survival probability
of the initial state $|0\rangle$ is more strengthened with larger
tunneling rate of the excited state. That is the essential meaning
of QZE of a system measured continuously.

Additionally, we can also investigate the relation of pulse and
continuous scheme in the future experiment. Theoretically, it could
be proved that if the effective decay times of the initial state in
the two schemes are the same, the interval between two sequential
measurements in pulse scheme and decay rate of the excited state in
continuous scheme should satisfy the relation \cite{L. S. Schulman}
of $\delta t\cdot\Gamma=4$. That is an important relationship
between the two schemes.


We have proposed two schemes to observe QZE in a superconducting
phase qubit: pulse and continuous measurement schemes. They are easy
and feasible for up-to-date technique. Our result show that QZE can
be demonstrated with the schemes  more clearly than   the trapped
ions experiment. Since QZE is essential in the implementation of
quantum information, the generalization of the proposed scheme to
other kinds of superconducting qubits \cite{Makhlin,Zhu} is
desirable.

\bigskip

We thank Prof. S.-L. Zhu for many helpful suggestion. This work was
supported by the NSFC (No. 11004065), the SKPBR of China (No.
2011CB922104), the NSF of Guangdong Province (No.
10451063101006312), and the Startup Foundation  of SCNU (No.
S53005).

\end{CJK*}
\end{document}